\documentclass[twocolumn, prl]{revtex4}

\usepackage{graphicx}% Include figure files
\usepackage{dcolumn}% Align table columns on decimal point
\usepackage{bm}% bold math

\begin{document}

\title{ A Renormalization Group Approach for Highly Anisotropic Fermion systems }

\author {S. Moukouri }

\affiliation{ Department of Physics and  Michigan Center for 
          Theoretical Physics \\
         University of Michigan, 2477 Randall Laboratory, Ann Arbor MI 48109}

\begin{abstract}
I apply a two-step density-matrix renormalization group method to the 
anisotropic two-dimensional tight-binding model. This study, which is a prelude
to the study of models of quasi-one dimensional materials, shows the
potential power of this approach for anisotropic fermionic models. I
find a ground-state energy which agrees with the exact value
up to four digits for systems as large as $24 \times 25$. 
This open new opportunities for simulations of fermions in two
dimensions.
\end{abstract}

\maketitle
Quasi-one dimensional organic \cite{reviewOC} and inorganic \cite{reviewIOC}
 materials have been the object
of an important theoretical interest for the last three decades.
The essential features of their phase diagram may be captured by 
the simple anisotropic Hubbard model (AHM),

\begin{eqnarray}
\nonumber H=-t_{\parallel}\sum_{i,l,\sigma}(c_{i,l,\sigma}^{\dagger}
c_{i+1,l,\sigma}+h.c.)\\
\nonumber -t_{\perp}\sum_{i,l,\sigma}(c_{i,l,\sigma}^{\dagger}c_{i,l+1,\sigma}+h.c.)+ \\ 
 U\sum_{i,l} n_{i,l,\uparrow}n_{i,l,\downarrow}+\mu \sum_{i,l,\sigma} 
n_{i,l,\sigma}.
\label{hamiltonian}
\end{eqnarray}
 
\noindent or a more general Hubbard-like model including longer range Coulomb 
interactions. 
For these highly anisotropic materials, $t_{\perp} \ll t_{\parallel}$.  
Over the years, the AHM has remained a formidable
challenge to condensed-matter theorists. Some important insights on this 
model or its low energy version, the g-ology model, have been obtained through 
the work of Bourbonnais and Caron \cite{bourbonnais,giamarchi} and others.
 They used a perturbative renormalization group approach to analyze the
crossover from 1D to 2D at low temperatures. More recently, Biermann et al.
\cite{bierman} applied the chain dynamical mean-field approach to study 
the crossover
from Luttinger liquid to Fermi liquid in this model. Despite this 
important progress, crucial information such as  the ground-state
 phase diagram, or most notably, whether the AHM 
 displays superconductivity, are still unknown. 
So far it has remained beyond the reach of numerical methods such as
the exact diagonalization (ED) or the quantum Monte Carlo (QMC) methods.
ED cannot  exceed lattices of about $4 \times 5$. It is likely to
remain so for many years unless there is a breakthrough in quantum
computations. The QMC method is plagued by the minus sign problem
and will not be helpful at low temperatures. The small value of $t_{\perp}$
implies that, in order to see the 2D behavior,  it will be necessary to reach 
lower temperatures than those usually studied for the isotropic 2D Hubbard 
model. Hence, even in the absence of the minus sign problem, in order to
work in this low temperature regime, the QMC algorithm requires special
stabilization schemes which lead to prohibitive cpu time \cite{white-qmc}.

I have shown in Ref.\cite{moukouri-TSDMRG} that this class of anisotropic 
models may  be studied using a two-step density-matrix renormalization
group (TSDMRG) method. The TSDMRG method is a perturbative approach
in which the standard 1D DMRG is applied twice.
In the first step, the usual 1D DMRG method \cite{white} is applied
to find a set of low
lying eigenvalues $\epsilon_n$ and eigenfunctions $|\phi_n \rangle$ of a
single chain. In the second step, the  2D Hamiltonian is then projected
onto the basis constructed from the tensor product of the $|\phi_n \rangle$'s.
This projection yields an effective one-dimensional Hamiltonian for
the 2D lattice,

\begin{eqnarray}
 \tilde{H} \approx \sum_{[n]} E_{\parallel [n]} |\Phi_{\parallel [n]}
\rangle \langle\Phi_{\parallel [n]}| -
 t_{\perp}\sum_{i,l,\sigma}(\tilde{c}_{i,l,\sigma}^{\dagger}\tilde{c}_{i,l+1}+h.c.)
\end{eqnarray}

\noindent where  $E_{\parallel [n]}$ is the sum of eigenvalues of the
different chains, $E_{\parallel[n]}=\sum_l{\epsilon_{n_l}}$;
$|\Phi_{\parallel [n]}\rangle$ are the corresponding eigenstates,
$|\Phi_{\parallel [n]}\rangle =  |\phi_{n_1}\rangle  |\phi_{n_2}\rangle ...
|\phi_{n_L} \rangle$; $\tilde{c}_{i,l}^{\dagger}$, and $\tilde{c}_{i,l}$
are the  renormalized matrix elements in the single chain basis. They are 
given by

\begin{eqnarray}
(\tilde{c}_{i,l}^{\dagger})^{n_l,m_l}=(-1)^{n_i}\langle \phi_{n_l}|{c}_{i,l}^{\dagger}
|\phi_{m_l}\rangle, \\
(\tilde{c}_{i,l})^{n_l,m_l}=(-1)^{n_i}\langle \phi_{n_l}|{c}_{i,l} |\phi_{m_l}\rangle, 
\end{eqnarray}

\noindent where $n_i$ represents the total number of fermions from sites $1$ to
$i-1$. For each chain, operators for all the sites are stored in a
single matrix

\begin{eqnarray}
\label{bm1}
\tilde{c}_{l}^{\dagger}=(\tilde{c}_{1,l}^{\dagger},...,
\tilde{c}_{L,l}^{\dagger}),\\
\label{bm2}
\tilde{c}_{l}=(\tilde{c}_{1,l},...,\tilde{c}_{L,l}).
\end{eqnarray}

\noindent Since the in-chain degrees of freedom have been integrated out,
the interchain couplings are between the block matrix operators in
Eq.(~\ref{bm1},~\ref{bm2}) which depend only on the chain index $l$.
In this matrix notation, the effective Hamiltonian is
one-dimensional and it is also studied by the DMRG method. The only
difference compare to  a normal 1D situation is that the local operators are now
$ms_2 \times ms_2$ matrices, where $ms_2$ is the number of states kept 
during the second step. 

The two-step method has previously been applied to anisotropic
two-dimensional Heisenberg models. In Ref.\cite{moukouri-TSDMRG2},
it was applied to the $t-J$ model but due to the absence  of an exact
result in certain limits, it was tested against ED results on small 
ladders only.
 A systematic analysis of its performance on a fermionic model on 2D lattices 
of various size has not been done. In this letter, as a prelude to the study of 
the AHM, I will apply the TSDMRG to the 
anisotropic tight-binding model on a 2D lattice, i.e., 
The AHM with $U=0$. 
I perform a comparison with the exact result of the tight-binding model. 
I was able to obtain agreement for the ground-state energies on the order of 
$10^{-4}$ for lattices of up to $24 \times 25$. I then discuss how these
calculations may extend to the interacting case. 

The tight-binding Hamiltonian is diagonal in the momentum space, 
the single particle energies are,

\begin{eqnarray} 
\epsilon_k=-2t_{\parallel} cos k_x-2t_{\perp}cos k_y-\mu, 
\end{eqnarray}

\noindent with $k=(k_x,k_y)$, $k_x=n_x\pi/(L_x+1)$ and $k_y=n_y\pi/(L_y+1)$ for
open boundary conditions (OBC); $L_x$, $L_y$ are respectively the linear
dimensions of the lattice in the parallel and transverse directions.. 
The ground-state
energy of an $N$ electron system is obtained by filling the lowest states 
up to the Fermi level,
$E_{[0]}(N)=\sum_{k < k_F}  \epsilon_k$. However in real space, this problem
is not trivial and it constitutes, for any real space method
such as the TSDMRG, a test having the same level of
difficulty as the case with $ U \neq 0$. This is because the term involving 
$U$ is diagonal in real space and the challenge of diagonalizing 
the AHM arises from the hopping term.

I will study the tight-binding model at quarter filling, $N/L_xL_y=1/2$, the
nominal density of the organic conductors known as the Bechgaard salts.
 Systems of up to $L_x\times L_y=L \times (L+1)=24 \times 25$ will be studied.
During the first step, I keep enough states ($ms_1$ is a few hundred)
so that the truncation error $\rho_1$ is less than $10^{-6}$. I
target the lowest state in each charge-spin sectors $N_x \pm 2,~ N_x \pm 1,~ N_x$
and $S_z \pm 1,~ S_z \pm 2$, $N_x$ is the number of electrons within the
chain. It is fixed such that $N_x/L_x=1/2$. There is a total of $22$ charge-spin states 
targeted at each iteration.

\begin{figure}
\includegraphics[width=3. in, height=2. in]{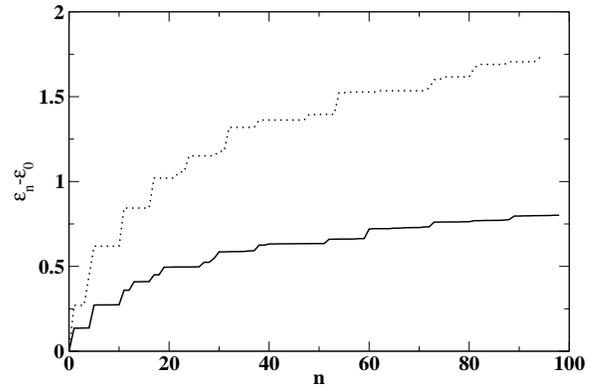}
\caption{Low-lying states of the 1D tight-binding model (full line) and
of the 1D Heisenberg spin chain (dotted line) for $L=16$ and $ms_2=96$. }
\vspace{0.5cm}
\label{density}
\end{figure}

For the tight-binding model, the chains remain disconnected if 
$t_{\perp} < \epsilon_0(N_x+1)-\epsilon_0(N_x)$ or $t_{\perp} < \epsilon_0(N_x)
-\epsilon_0(N_x-1)$, where $N_x$ is the number of electons on single chain.
 In order to observe transverse motion, it is necessary that at least 
$t_{\perp} \agt \epsilon_0(N_x+1)-\epsilon_0(N_x)$ and $t_{\perp} \agt \epsilon_0(N_x) -\epsilon_0(N_x-1)$. These two conditions are satisfied only if $\mu$ is 
appropietly chosen.
 The values listed in Table (\ref{param}) corresponds to 
$\mu= (\epsilon_0(N_x+1)-\epsilon_0(N_x-1))/2$. This treshold varies with $L$. I give
in Table (\ref{param}) the values of $t_{\perp}$ chosen for different
lattice sizes. In principle, for the TSDMRG to be accurate, it is necessary 
that $\Delta \epsilon=\epsilon_{n_c}-\epsilon_{0}$, where $\epsilon_{n_c}$ is 
the cut-off, be such that $\Delta \epsilon/t_{\perp} \gg 1$.
But in practice, I find that I can achieve accuracy up to the fourth 
digit even if $\Delta \epsilon/t_{\perp} \approx 5$ using the finite system
method. Five sweeps were necessary to reach convergence. Note that this 
conclusion is somewhat different from my earlier estimate of 
$\Delta \epsilon/t_{\perp} \approx 10$ for spin systems \cite{moukouri-TSDMRG2}. This 
is because in Ref.(\cite{moukouri-TSDMRG2}), I used the infinite system method 
during the second step.  

The ultimate success of the TSDMRG depends on the density of the
low-lying states in the 1D model. For fixed $ms_2$ and $L$, it is, for 
instance, easier to reach larger $\Delta \epsilon/J_{\perp}$ in the anisotropic spin
one-half Heisenberg model, studied in  Ref.\cite{moukouri-TSDMRG},
than  $\Delta \epsilon/t_{\perp}$ for the tight-binding model as shown 
in Fig.\ref{density}. For $L=16$, $ms_2=96$, and $J_{\perp}=t_{\perp}=0.15$,
I find that $\Delta \epsilon/J_{\perp} \approx 10$, 
while $\Delta \epsilon/t_{\perp} \approx 5$. Hence, the TSDMRG method will be more 
accurate for a spin model than for the tight-binding model. Using the infinite
system method during the second step on the anisotropic Heisenberg model
with $J_{\perp}=0.1$, I can now reach an agreement of about
$10^{-6}$ with the stochastic QMC method.

\begin{table}
\begin{ruledtabular}
\begin{tabular}{cccc}
  & $8 \times 9$ & $16 \times 17$ & $24 \times 25$ \\
\hline
 $t_{\perp}$ &-0.28 & -0.15& -0.1  \\
 $\mu$ & -1.2660 & -1.3411& -1.3657 \\
$\Delta \epsilon/t_{\perp}$ & 6.42 & 5.40 & 5.78 \\ 
\end{tabular}
\end{ruledtabular}
\caption{Transverse hopping and chemical potential used in the 
simulations for different lattice sizes}
\label{param}
\end{table}

Two possible sources of error can contribute to reduce the accuracy
in the TSDMRG with respect to the conventional DMRG. They are  the
truncation of the superblock from $4 \times ms_1$ states to only $ms_2$
states and the use of three blocks instead of four during the second
step. In Table (\ref{eg1}) I analyze the impact of the reduction of
the number of states to $ms_2$ for three-leg ladders. The choice of 
three-leg ladders is motivated by the fact that at this point, the
TSDMRG is equivalent to the exact diagonalization of three reduced
superblocks. It can be seen that as far as 
$t_{\perp} \agt \epsilon_0(N_x+1)-\epsilon_0(N_x)$ and
 $t_{\perp} \agt \epsilon_0(N_x) -\epsilon_0(N_x-1)$, 
  the TSDMRG at this point
is as accurate as the 1D DMRG. Note that
the accuracy remains nearly the same irrespective of $L$ as far as the 
ratio $\Delta \epsilon/t_{\perp}$ remains nearly constant.
Since $\Delta \epsilon$ decreases when $L$ increases, $t_{\perp}$ must be
decreased in order to keep the same level of accuracy for fixed
$ms_2$. In principle, following this prescription, much larger
systems may be studied. $\Delta \epsilon/t_{\perp}$ does not have to be very 
large, in this case it is about $5$, to obtain very good agreement with
the  exact result.

\begin{table}
\begin{ruledtabular}
\begin{tabular}{cccc}
 $ms_2$ & $8 \times 3$ & $16 \times 3$ & $24 \times 3$ \\
\hline
 $64$ & -0.241524 & -0.211929 & 0.204040 \\
 Exact  & -0.241524&  -0.211931 & 0.204049 \\
\end{tabular}
\end{ruledtabular}
\caption{Ground-state energies of three-leg ladders.}
\label{eg1}
\end{table}

The second source of error is related to the fact that   the
effective single site during the second step is now a chain
having $ms_2$ states, I am thus   forced to use three blocks instead 
of four to reduce the computational burden. In Table (\ref{eg2}),
it can be seen that this results in a reduction in accuracy of 
about two orders of magnitude with respect to those of three 
leg-ladders. These results are nevertheless very good given the 
relatively modest computer power involved. All calculations were
done on a workstation.

\begin{table}
\begin{ruledtabular}
\begin{tabular}{cccc}
 $ms_2$ & $8 \times 9$ & $16 \times 17$ & $24 \times 25$ \\
\hline
 $64$ & -0.24761 & -0.21401 & 0.20504 \\
 $100$ & -0.24819 & -0.21414 & 0.20509\\
$120$& -0.24832 & -0.21419 & \\
 Exact  & -0.24857& -0.21432 & 0.20519\\
\end{tabular}
\end{ruledtabular}
\caption{Ground-state energies for different lattice sizes; 
a  single  state was targeted in the second step.}
\label{eg2}
\end{table}

The DMRG is less accurate when three blocks are used instead of four.
This can be understood  by applying the following view on the formation
of the reduced density matrix. The construction of the reduced density matrix
may be regarded as a linear mapping $u_\Psi:~ {\bf F^*} 
\rightarrow {\bf E}$, where ${\bf E}$ is the system, ${\bf F}$ is the 
environment and, ${\bf F^*}$ is the dual space of ${\bf F}$. 
Using the decomposition
of the superblock wave function $\Psi_{[0]}= \sum_i \phi_i^L \otimes \phi_i^R$,
with $\phi_i^L \in {\bf E}$ and $\Phi_i^R \in {\bf F}$, 
for any $\phi^* \in F^*$, 

\begin{eqnarray}
u_\Psi(\phi^*)=\sum_{i=1} \langle\phi^*|\phi_i^R\rangle\phi_i^L.
\end{eqnarray}

\noindent Let $|k\rangle,~k=1,...dim{\bf E}$ and $|l\rangle,~l=1,...dim{\bf F}$
be the basis of ${\bf E}$ and ${\bf F}$ respectively. Then, $|l\rangle$ has 
a dual basis $\langle l^*|$ such that $\langle l^*|l\rangle=\delta_{l,l^*}$.
The matrix elements of $u_\Psi$ in this basis are just the coordinates of
the superblock wave funtion $\Phi_{[0]_{k,l}}$. The rank $r$ of this mapping,
which is also the rank of the reduced density matrix is
always smaller or equal to the smallest rank of ${\bf E}$ and 
${\bf F}$, $r < Min(dim {\bf E}, dim {\bf F})$. Hence, if $ms_2$ states are 
kept in the two external blocks, the number of non-zero eigenvalues of $\rho$ 
cannot be larger than $ms_2$. Consequently, some states which have 
non-zero eigenvalues in the normal four block configuration will be missing.
 A possible cure to this problem
is to target additional low-lying states above $\Psi_{[0]}(N)$. The weight of
these states in $\rho$ must be small so that their role is simply to add
the missing states not to be described accurately themselves. A larger weight 
on these additional states would  lead to the reduction of the accuracy for a 
fixed $ms_2$. In table (\ref{eg3}), I show the improved energies when, 
besides the ground state, I target the lowest states of
the spin sectors $S_z=-1$ and $S_z=+1$ with $N$ electrons. The weights were 
respectively
$0.995$, $0.0025$, and $0.0025$ for the three states.  This lowers $E_{[0]}(N)$
in all cases, but the gain does not appear to be spectacular. But I do
not know whether this is due to my choice of perturbation of $\rho$ 
or whether even the algorithm with four blocks would not yield better
$E_{[0]}(N)$. If the lowest sectors with $N+1$ and $N-1$ electrons which
have $S_z=\pm 0.5$ are projected instead, I find that the results are 
similar to those with $S_z=\pm 1$ sectors, there are possibly many ways
to add the missing states. A more systematic approach to this problem has 
recently been suggested \cite{white2}. It is based on using a local
perturbation to build a correction to the density matrix from the site at the 
edge of the system. Here, such a perturbation would be $\Delta \rho= \alpha
c_l^{\dagger} \rho c_l$, where $\alpha$ is a constant, 
$\alpha \approx 10^{-3}-10^{-2}$, and $c_l^{\dagger},~c_l$ are the creation
and annihilation operators of the chain at the edge of the system.
This type of perturbation resulted in an accuracy gain of more than
an order of magnitude in the case of a spin chain \cite{white2}. The three
block method was found to be on par with the four block method. It will
be interesting to see in a future study how this type of local perturbation
performs within the TSDMRG.    

\begin{table}
\begin{ruledtabular}
\begin{tabular}{ccc}
 $ms_2$ & $8 \times 9$ & $16 \times 17$  \\
\hline
 $64$ & -0.24803 & -0.21401  \\
 $100$ & -0.24828 & -0.21417 \\
 Exact  & -0.24857&  -0.21432\\
\end{tabular}
\end{ruledtabular}
\caption{Ground-state energies for different lattice sizes; 
three states were targeted in the second step: the ground state
itself and the lowest states of $S_z=0$ and $S_z=1$ sectors.}
\label{eg3}
\end{table}

\begin{figure}
\includegraphics[width=3. in, height=2. in]{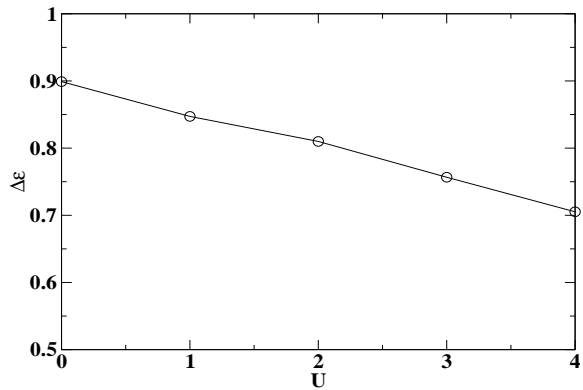}
\caption{Width $\Delta \epsilon$ for the low-lying states of the 1D Hubbard chain
as function of $U$ for  $L=16$ and $ms_2=128$. }
\vspace{0.5cm}
\label{rolu}
\end{figure}

To conclude, as a first step to the investigation of interacting electron 
models, I have shown that the TSDMRG can successfuly be applied to the  
tight-binding model. The agreement with the exact result is very good and can 
be improved since the computational power involved in this study
was modest. The extension to the AHM with $U \neq 0$
is straightforward. There is no additional change in the algorithm 
since the term involving $U$ is local and thus treated during the
1D part of the TSDMRG. The role of $U$ is to reduce $\Delta \epsilon$ as 
shown in Fig.\ref{rolu}. For fixed $L$ and $ms_2$, $\Delta \epsilon$ decreases
linearly with increasing $U$. For $L=16$ and $ms_2=128$, I anticipate that
for $U \alt 3$ the interacting system results will be on the same level or
better than those with $ms_2=100$ for the same value of $L$.

\begin{acknowledgments}
I wish to thank A.M.-S. Tremblay for very helpful discussions. I thank K.L.
Graham for reading the manuscript. This work was supported by the NSF Grant No. DMR-0426775.

\end{acknowledgments}

\end{document}